\documentstyle[epsfig,11pt]{article}
\newcommand{\be}{\begin{equation}}
\newcommand{\ee}{\end{equation}}
\newcommand{\ben}{\begin{eqnarray}}
\newcommand{\een}{\end{eqnarray}}
\newcommand{\bc}{\begin{center}}
\newcommand{\ec}{\end{center}}

\begin{document}

\title{On the possibility of an astronomical detection of chromaticity effects in
microlensing by wormhole-like objects}
\author{Diego F.
Torres$^{1,}$\thanks{e-mail: dtorres@princeton.edu}, Ernesto F.
Eiroa$^{2,}$\thanks{e-mail: eiroa@iafe.uba.ar}, and Gustavo E.
Romero$^{3,}$\thanks{e-mail: romero@irma.iar.unlp.edu.ar. Member
of CONICET}
\\{\small $^1$ Physics Department, Princeton University,
NJ 08544, USA}\\ {\small $^2$ Instituto de Astronom\'{\i}a y
F\'{\i}sica del Espacio, C.C. 67, Suc. 28, 1428, Buenos Aires,
Argentina}\\ {\small $^3$Instituto Argentino de
Radioastronom\'{\i}a, C.C.5, 1894 Villa Elisa, Buenos Aires,
Argentina} }

\maketitle

\begin{abstract}
We study the colour changes induced by blending in a wormhole-like
microlensing scenario with extended sources. The results are
compared with those obtained for limb darkening. We assess the
possibility of an actual detection of the colour curve using the
difference image analysis method.

\end{abstract}

\section{Introduction}

This paper continues our study on the possible observational
effects that struts of negative masses would produce if they are
isolated in space \cite{our}. Since wormhole structures require
the violation of some of the most sensitive energy conditions at
the wormhole throat, wormholes are natural candidates --if they
exist at all-- for stellar size negative mass objects. Different
wormhole solutions have been presented in the literature after the
leading work of Morris and Thorne \cite{motho} (see for example
Refs. \cite{wh}). Many of these solutions actually present a
negative energy density and open the possibility of having a total
negative mass. However, only a few works deal with the problem of
developing observational tests for the existence of wormhole-like
objects. Our aim in the present series of papers is to turn the
speculation on macroscopic amounts of negative masses into an
experimental question, one whose answer could be reached by
current astrophysical observations.\\

In a recent paper \cite{eiroa-croma}, we have studied the
gravitational microlensing scenario that a negative mass point
lens would produce over an extended source. This allowed us to
present more realistic light curves for wormhole microlensing
events than those obtained earlier by Cramer et al. \cite{cramer}.
Using the formalism introduced in Refs. \cite{Han}, we computed
the effects of a finite source extent on the spectral features of
microlensing. We showed that limb darkening of the intensity
distribution on a stellar source induces specific chromaticity
effects that are very different from what is expected in the
positive mass lens case. The possibility of using multi-colour
optical observations to search for galactic or inter-galactic
natural wormhole-like objects was then foreseen.\\

Detection of the extended source effects from colour measurements,
instead of single band photometry, is interesting because of two
facts (see Ref. \cite{Han2} for further discussion). Firstly, by
detecting the colour curves the extended nature of the source is
revealed: if the source approaches very close to the lens caustics
but do not cross them, the induced amplification can always be
mimicked by changes in the lensing parameters of a point-like
object. By contrast, the colour curves can not be mimicked by any
such changes: a point source lensing event should always be
achromatic. Secondly, the colour curve allows one to measure the
lens proper motion quite easily, without the need of fitting the
entire light curve. \\

However, measurements of the colour curve can actually be hampered
by light blending caused from nearby and background sources, which
also causes chromaticity effects. Han et al. \cite{Han} have
demonstrated that even for a small fraction (less than 2\%) of
blended light, the colour changes caused by blending can be
equivalent in magnitude to those caused by limb-darkening.
Therefore, in order to get predictions for a colour curve, it is
essential to take blending into account, and to remove, somehow,
its effects.\\

In the present letter we shall analyze the chromaticity effects
produced, in the case of a wormhole-like microlensing event, by
blending of other stars. In addition, we shall estimate the
likelihood of carrying out an actual observation of the colour
curves using the difference image analysis method within current
technological capabilities.

\section{A brief summary of microlensing formulae}

The amplification produced by gravitational lensing of a point
source is given by \cite{cramer}
\begin{equation}
A_{0}=\frac{B_{0}^{2}\pm 2}{B_{0}\sqrt{B_{0}^{2}\pm 4}}  \label{1}
,
\end{equation}
where the plus sign corresponds to positive mass and the minus
sign to negative mass lensing, and $B_{0}=b_{0}/R_{E}$ is the
lens-source separation in units of the Einstein radius $R_{E}$,
\begin{equation}
R_{E}=\sqrt{\frac{4G|M|}{c^{2}}\frac{D_{ol}D_{ls}}{D_{os}}}.
\label{2}
\end{equation}
As usual, $D_{os}$ is the observer-source distance, $D_{ol}$ is
the observer-lens distance, $D_{ls}$  is the lens-source distance,
and $M$ the mass of the gravitational lens. For an extended
circular source, instead, the light curve is given by
\cite{eiroa-croma}
\begin{equation}
A=\frac{\int_{0}^{2\pi }\int_{0}^{r_{*}}{\mathcal I}(r,\varphi
)A_{0}(r,\varphi )rdrd\varphi }{\int_{0}^{2\pi
}\int_{0}^{r_{*}}{\mathcal I} (r,\varphi )rdrd\varphi } .
\label{3}
\end{equation}
Here, $(r,\varphi )$ are polar coordinates in a reference frame
placed in the center of the star, $r_{*}$ is the radius of the
source, and $ {\mathcal I}(r,\varphi )$ is its surface intensity
distribution. For a radially symmetric distribution, the previous
expression transforms into (defining the dimensionless radius
$R=r/R_{E}$)
\begin{equation}
A=\frac{\int_{0}^{2\pi }\int_{0}^{R_{*}}{\mathcal
I}(R)A_{0}(R,\varphi )RdRd\varphi }{2\pi \int_{0}^{R_{*}}{\mathcal
I}(R)RdR} , \label{5}
\end{equation}
where $R_{*}=r_{*}/R_{E}$ is the dimensionless radius of the star.
If the lens is moving with constant velocity $v$, the lens-source
separation (in units of the Einstein radius) is given by
\begin{equation}
B(T)=\frac{b(t)}{R_{E}}=\sqrt{\left( B_{0}+R\sin \varphi \right)
^{2}+\left( -T+R\cos \varphi \right) ^{2}},  \label{7}
\end{equation}
where $T=vt/R_{E}$ (see Ref. \cite{eiroa-croma} for a helpful plot
and further details). Replacing $B_{0}$ in Eq. (\ref{1}) by its
time-dependent partner, $B(T)$, and using this expression in Eq.
(\ref{5}), we arrive, for a given intensity distribution
${\mathcal I}(R)$, at the light curves produced by lensing in an
extended source case.

\section{Blending}

The obscuration of the intensity profile of a star towards its
border is known as limb darkening. An extended source microlensing
event become chromatic as a consequence of this effect, see for
instance Ref. \cite{Han}. The colour change caused by limb
darkening of the source star can be computed using \cite{Han}
\begin{equation}
\Delta (m_{\nu _{2}}-m_{\nu _{1}})=-2.5\log \left( \frac{A_{\nu
_{2}}}{ A_{\nu _{1}}}\right) , \label{9}
\end{equation}
where $A_{\nu _{1}}$ and $A_{\nu _{2}}$ are the amplifications in
two different wavelength bands, $\nu _{1}$ and $\nu _{2}$. For the
intensity profile we shall take, in terms of the  radius $R$, and
as in Refs. \cite{eiroa-croma,Han},
\begin{equation}
{\mathcal I}_{\nu }(R)=1-C_{\nu }\left( 1-\sqrt{1-\left(
\frac{R}{R_{*}} \right) ^{2}}\right) , \label{11}
\end{equation}
with the limb-darkening coefficients $C_{\nu _{1}}=0.503$ ,
$C_{\nu _{2}}=1.050$ corresponding to the I and U bands of a
K-giant with $ T_{{\rm eff}}=4750K$.\\

But the light curve of a microlensing event can also be chromatic
by another effect: blending. Basically, the light flux of a source
star can be affected by blended light of other unresolved stars,
having themselves different colours, what results in a change of
the colour curve. If we consider both effects at the same time,
limb darkening and blending, the so generated colour curve was
recently computed by Han et al. \cite{Han2} to be
\begin{equation}
\Delta (m_{\nu _{2}}-m_{\nu _{1}})=-2.5\log \left[ \left(
\frac{A_{\nu _{2}}+f_{\nu _{2}}}{A_{\nu _{1}}+f_{\nu _{1}}}\right)
\left( \frac{1+f_{\nu _{2}}}{1+f_{\nu _{1}}} \right) ^{-1}
\right], \label{12}
\end{equation}
where $f_{\nu _{i}}$ are the fractions of the blended light in the
individual wavelength bands. These fractions depend on the
specific situation and will be different for different
backgrounds. The colour changes due only to the blending effect
will be the difference between the colour curve of the limb
darkening event affected by blending (Eq. \ref{12}) and the colour
curve for limb darkening alone (given in Eq. (\ref{9}) and Ref.
\cite{eiroa-croma}). To ease the comparison with the standard
(i.e. possitive mass) case, we shall adopt the blending
coefficients as $f_{U}=1.39\%$ and $f_{I}=1.04\%$, and a source
star with radius $R_{*}=0.1$ \cite{Han2}. Our new results,
including the effect of blending, and for different impact
parameters $k=b_{0}/r_{*}=B_{0}/R_{*}$, are shown in Figures 1 and
2.\\

\begin{figure}[h!]
\begin{center}
\includegraphics[width=8cm,height=9cm]{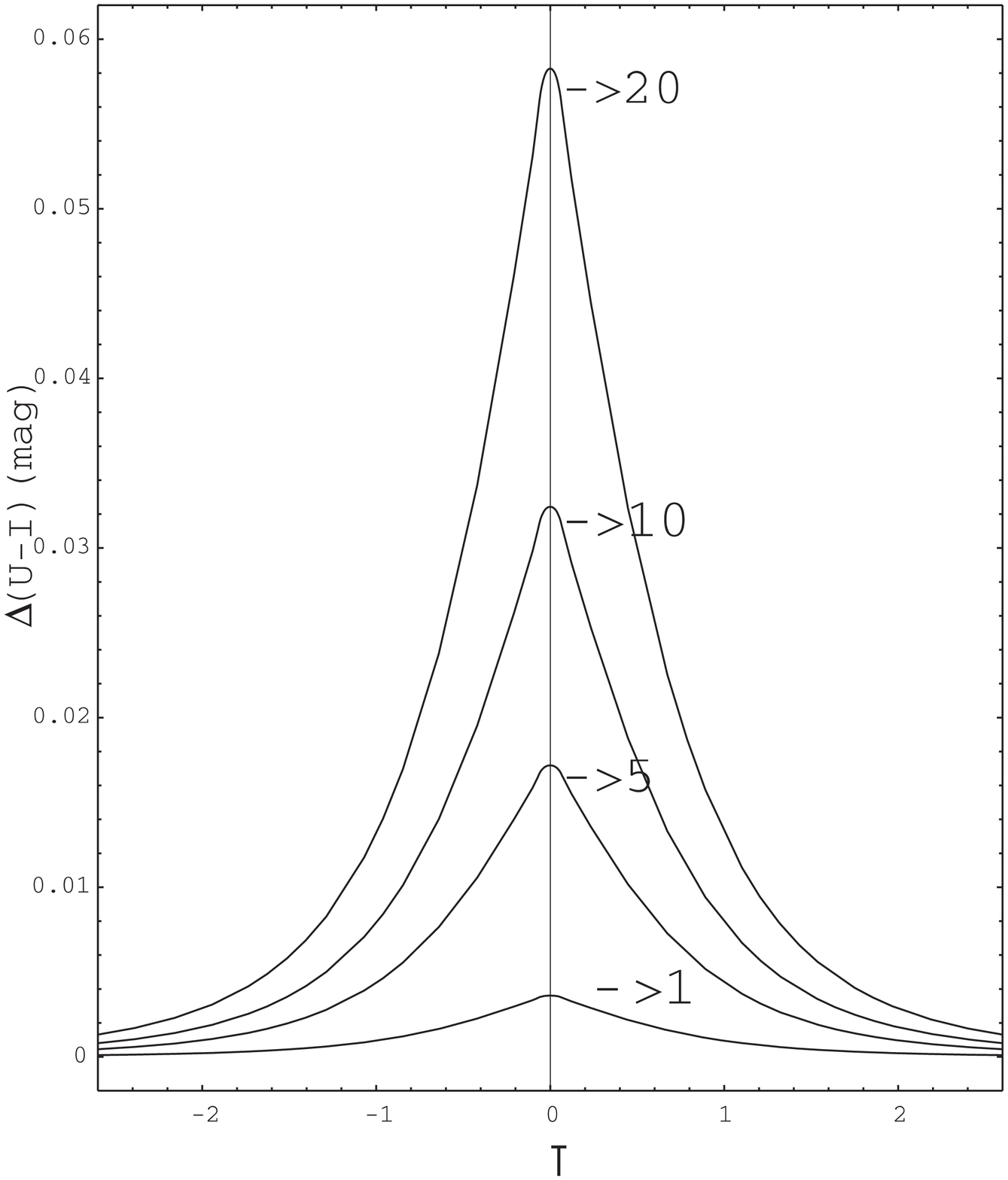}
\hspace{0.3cm}
\includegraphics[width=8cm,height=9cm]{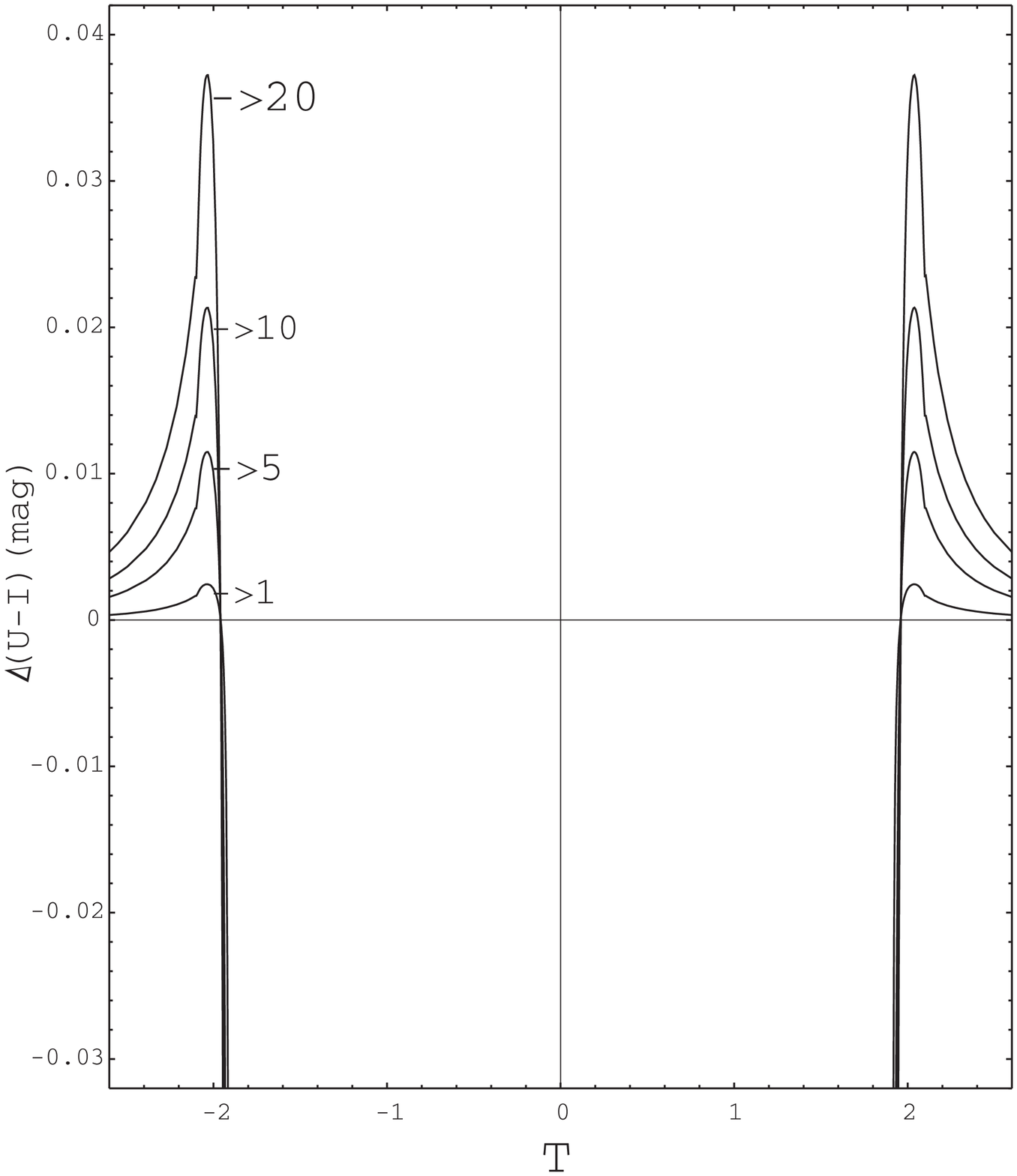}\\
\vspace{0.3cm}
\includegraphics[width=8cm,height=9cm]{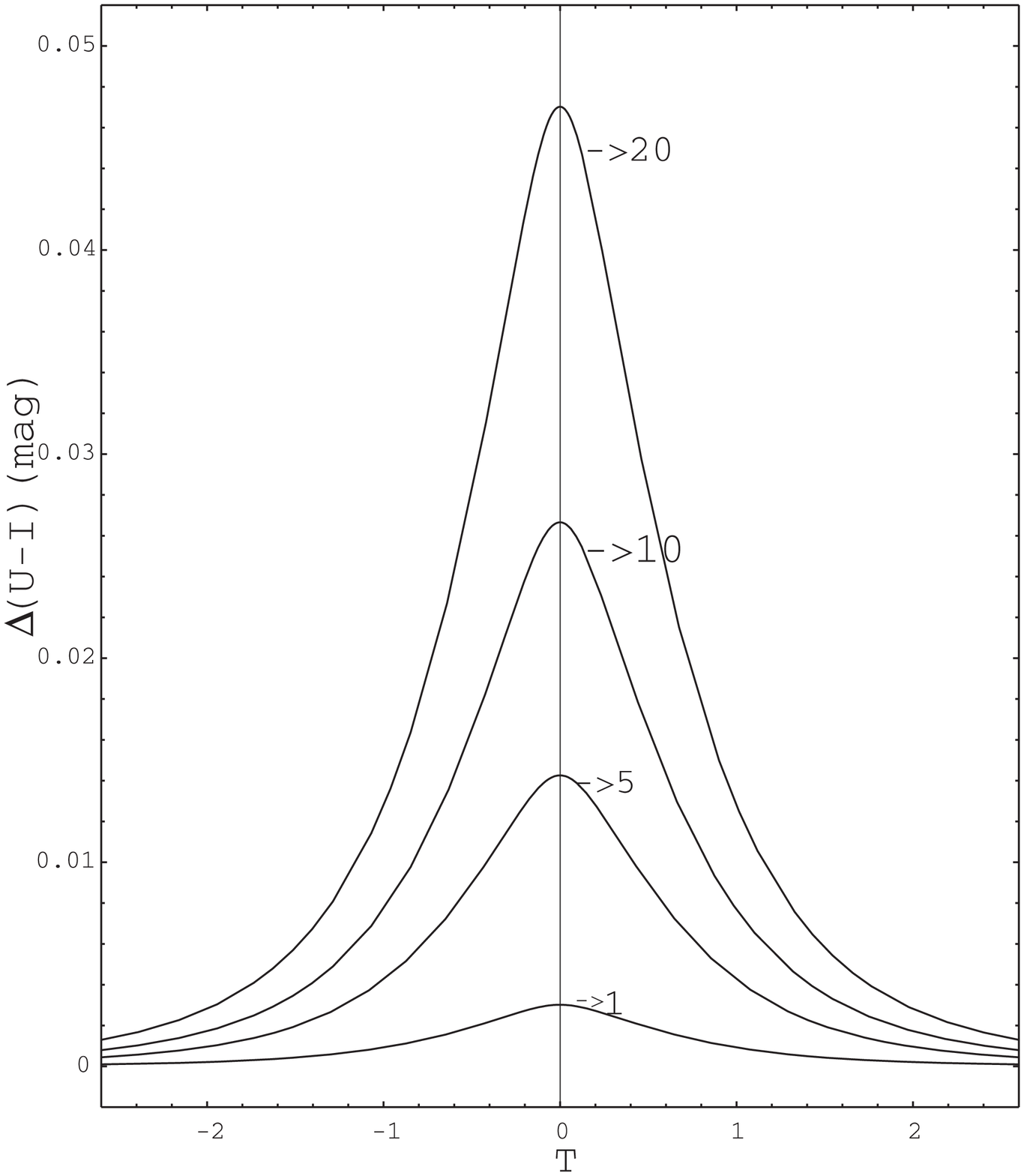}
\hspace{0.3cm}
\includegraphics[width=8cm,height=9cm]{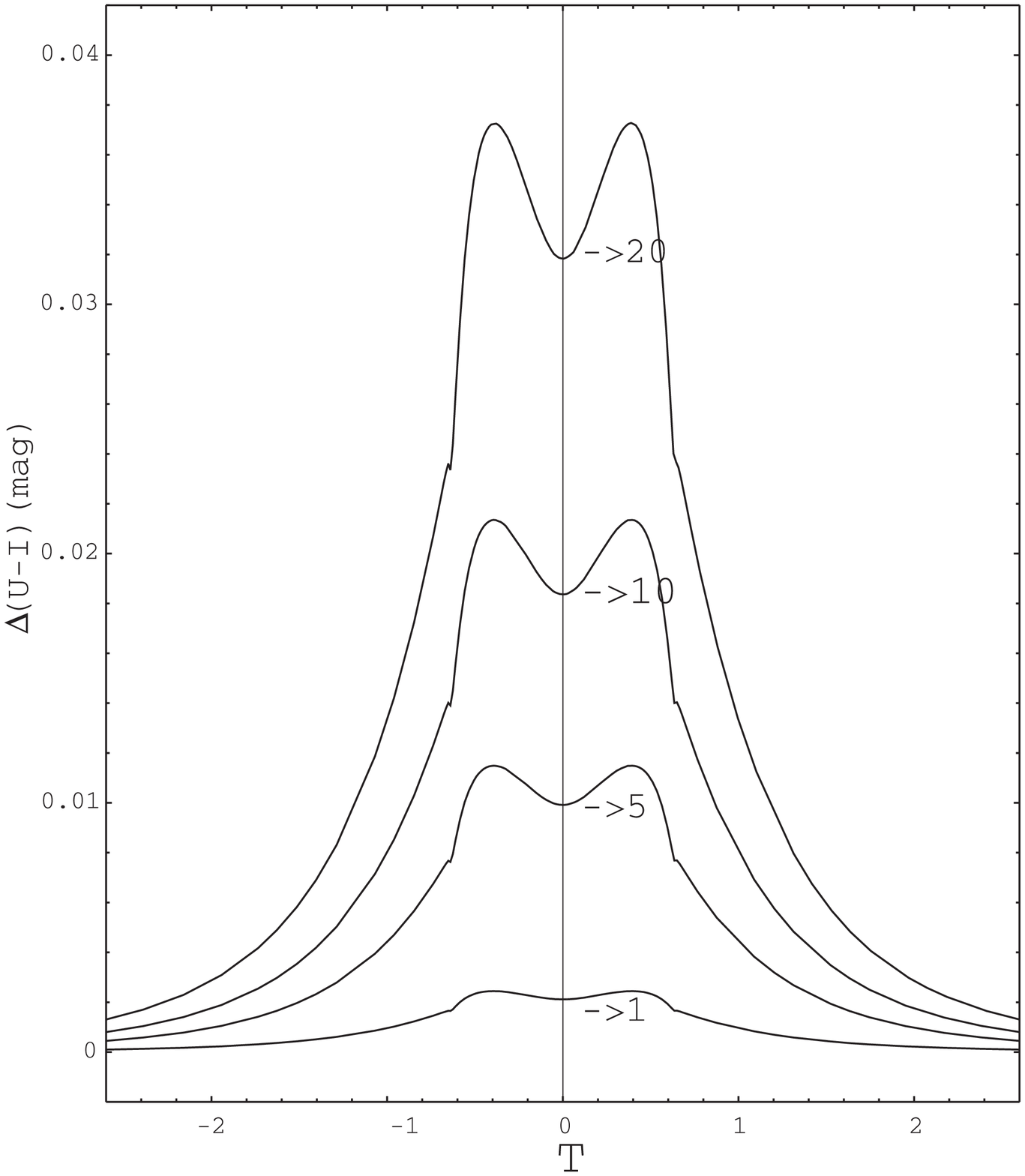}
 \caption{Colour changes produced by blending in microlensing events.
 The number of blended stars is shown in each
curve. Upper left: Positive lensing, impact parameter
$k=b_{0}/r_{*}=0.5$. Upper right: Negative lensing, same impact
parameter. Bottom Left: Positive lensing,
 $k=2$. Bottom right: Negative lensing, $k=20$. The
 source star has dimensionless radius $R_{*}=0.1$.
\label{1}}
\end{center}
\end{figure}

\begin{figure}
\begin{center}
\includegraphics[width=8cm,height=9cm]{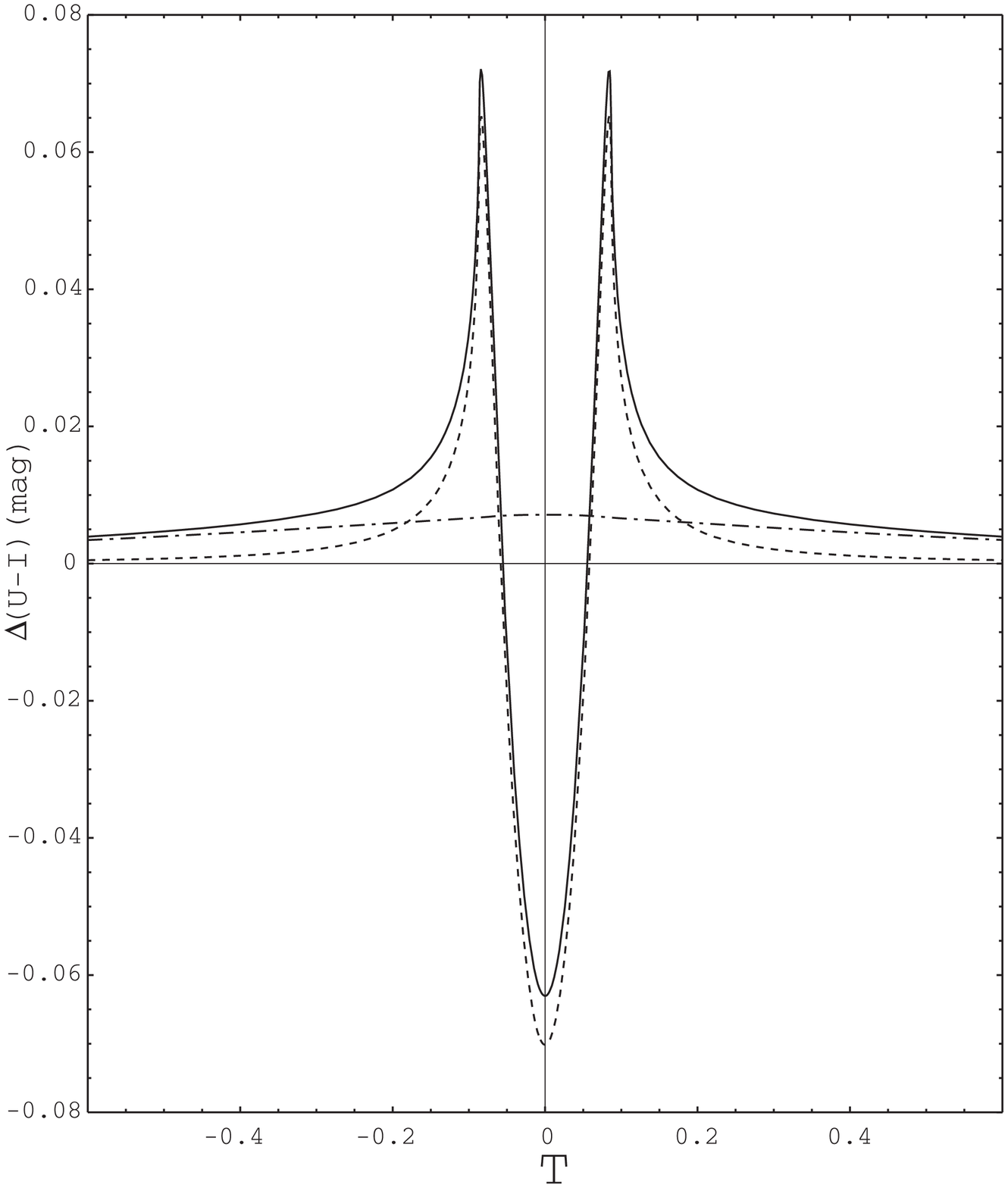}
\hspace{0.3cm}
\includegraphics[width=8cm,height=9cm]{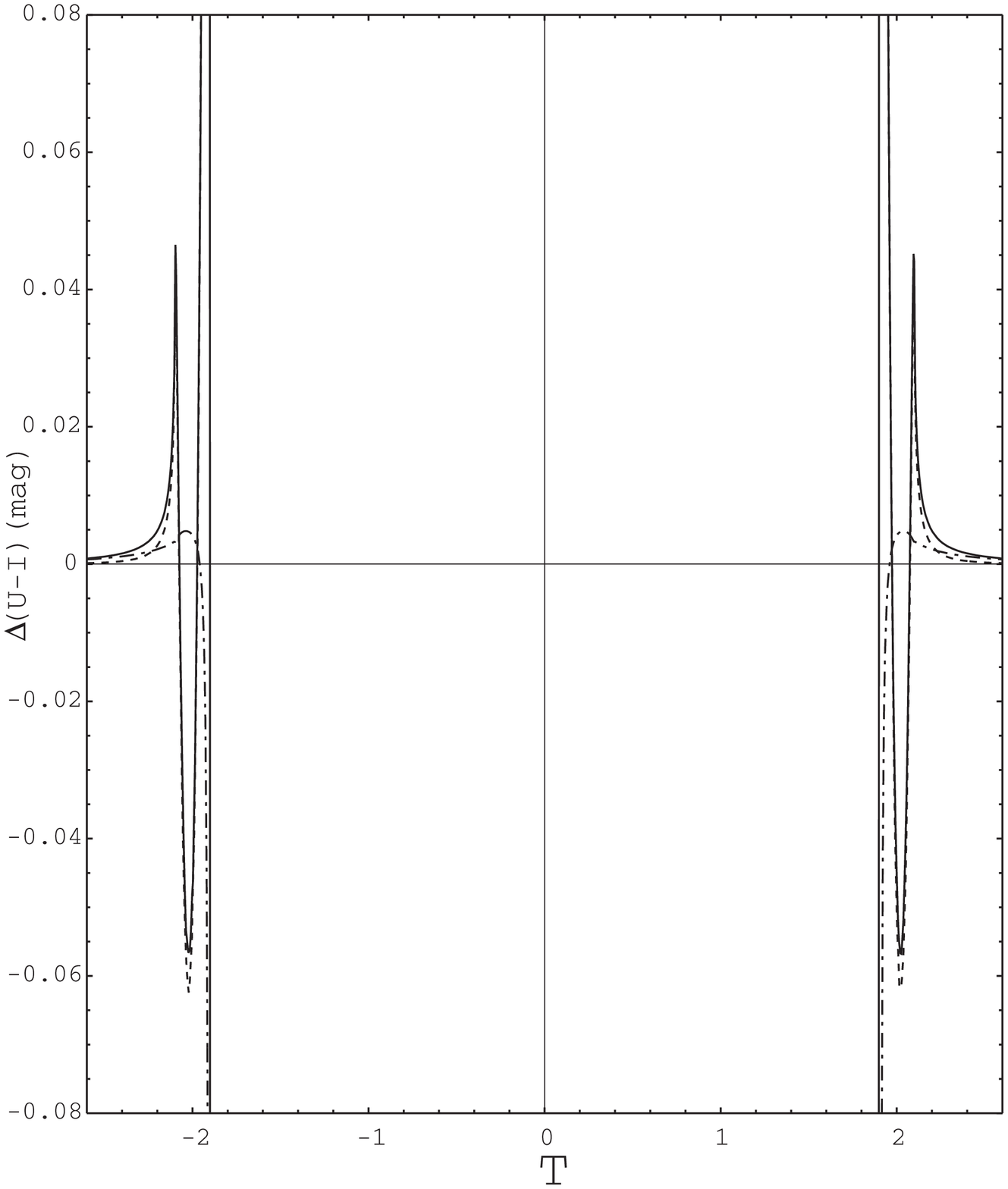}\\
\vspace{0.3cm}
\includegraphics[width=8cm,height=9cm]{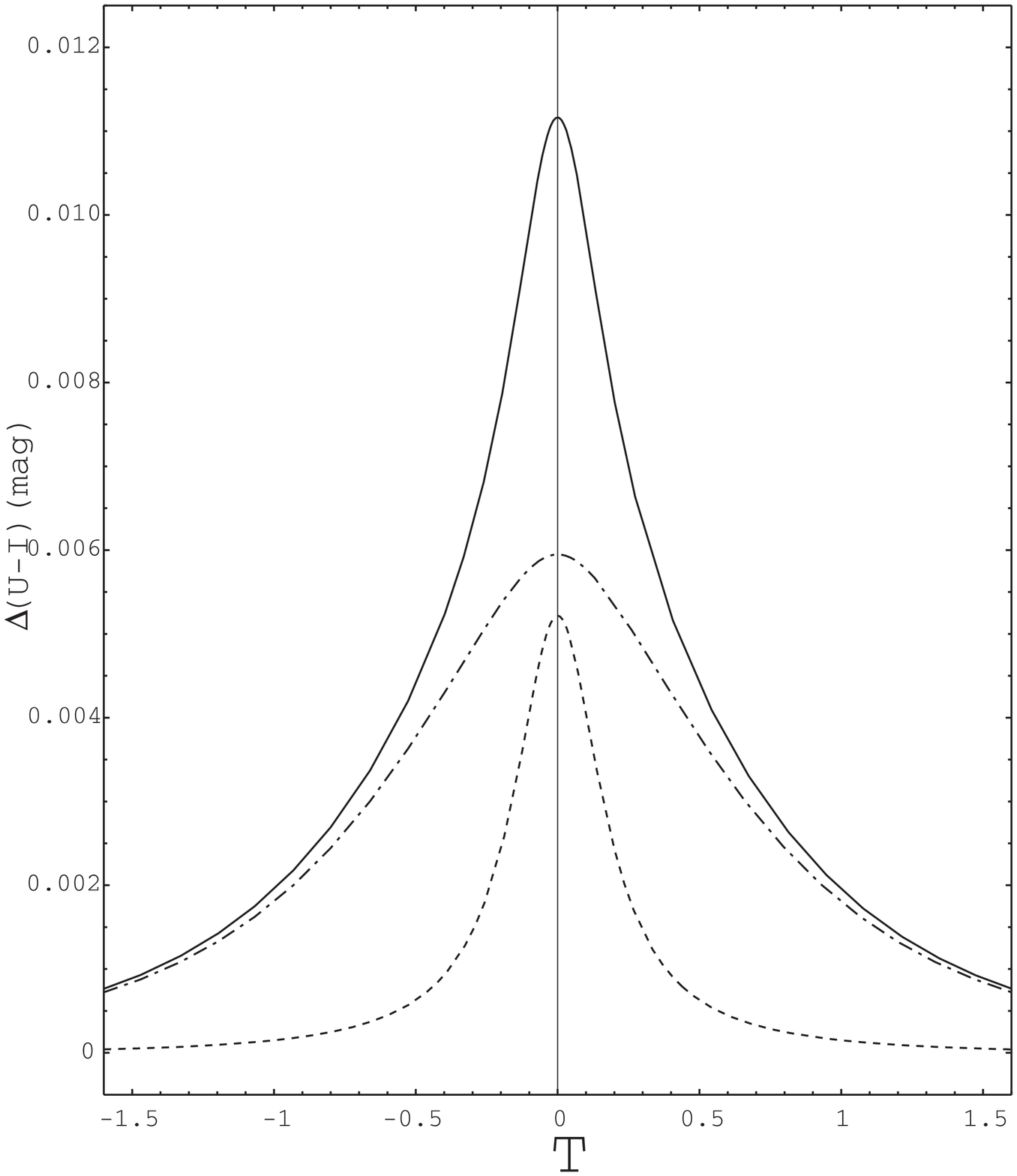}
\hspace{0.3cm}
\includegraphics[width=8cm,height=9cm]{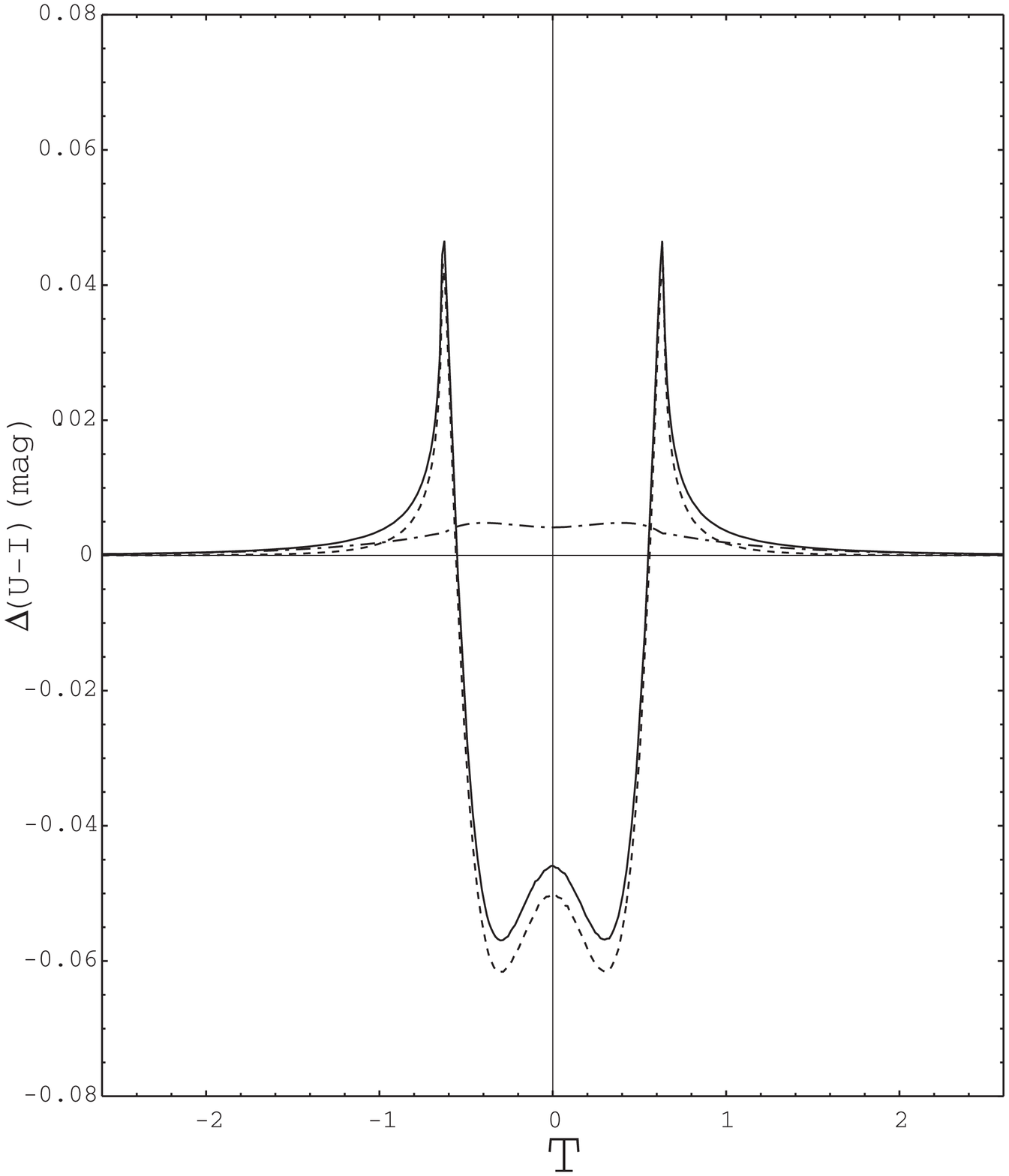}
 \caption{Full and partial contributions to chromaticity effects in
 microlensing.
 The solid line is the full colour curve of an event affected by blending
 (two blended stars), the dash curve represents the colour curve if the same event
 is not affected by blending, whereas the dash-dot curve gives only the colour changes due
 to the blending effect. Upper Left:
  Positive lensing, impact parameter $k=b_{0}/r_{*}=0.5$.
 Upper right: Negative lensing, same impact
parameter. The umbra is no longer present, but rather there is a
plateau ($ \Delta (U-I) \approx -0.31$), that dominates
 the umbra region in the
colour curves, produced only by the blended light. This plateau is
not shown in the figure in order to show the detail in the upper
portions of the colour curve. Bottom Left: Positive lensing,
 $k=2$. Bottom right: Negative lensing, $k=20$.
The
 source star has dimensionless radius $R_{*}=0.1$. \label{2}}
\end{center}
\end{figure}

The colour curves without blending present an umbra region in the
negative lensing case (no light reaches the observer) when the
impact parameter is small ($k<20$) \cite{eiroa-croma}. Considering
blending, instead,  we now discover that this umbra is no longer
present, but rather that there is a `plateau' ($ \Delta (U-I)
\approx -0.31$) in the colour curves, produced only by the blended
light. This plateau is not directly shown in the figures in order
to show the detail in the upper portion of the colour curve. This
new effect has important implications in the full colour curves,
as Figure 2 shows.\\

In the case of an ordinary lens, the colour curves affected by
blending are very similar to the photometric ones, see
\cite{eiroa-croma} for a comparison. We see that as it gets closer
to the star, the color of the observed source becomes redder due
to the differential amplification of the coldest regions. When the
lens transits towards the star interior, the hot center starts to
dominate the amplification, producing a dramatic change in the
slope. \\

For the limb darkening colour curve (dash curve in Figure 2), the
spectral changes start long before than in the standard situation.
Initially, the source also becomes redder and then experiences a
switch when shorter wavelengths begin to dominate. Contrary to
what happens with positive masses, the spectral trend changes
again, with the source appearing colder and colder until it
vanishes in the umbra during the transit. When the source is seen
again, the inverse behaviour is observed. If we now take into
account the blending effect as well, the existence of the
previously mentioned plateau, instead of the umbra region, make
the colour curve to change its trend again, towards the blue
region. During the transit, it is the blended contribution the one
that dominates the colour curve. It makes sense: blending fluxes
come from stars whose light is not deflected by the wormhole-like
object, and so the typical umbra effect is absent. Blending, then,
and contrary to the positive mass case (where the pattern of the
colour curve is maintained with only slight changes in the actual
values for $\Delta (U-I)$), noticeably affects the form of the
colour curve in a wormhole-like microlensing event.\\

The difference between the negative and the positive colour curves
(that we show for comparison in the same set of figures) continues
to be very clear, and hence, these combined effects allow to
distinguish between the different kind of lenses. We shall now
focus on demonstrating that the colour curve can actually be
observed with current technology in typical cases.

\section{The DIA colour curve }

The difference image analysis (DIA) is a method to measure
blending-free light colour variations by subtracting an observed
image from a convolved and normalized reference one. The flux
would then be, within DIA, \be F_\nu=F_{\nu,{\rm obs}}-F_{\nu,{\rm
ref}}=(A_\nu-1)F_{\nu,0},\ee where $F_{\nu,{\rm obs}}=A_\nu
F_{\nu,0}+B_\nu$ and $F_{\nu,{\rm ref}}=F_{\nu,0}+B_\nu$ stand for
the source star fluxes measured from the images obtained during
the progress of the microlensing event, and from the reference
(unlensed) image, respectively. $B_\nu$ is the blended flux. Then,
the DIA colour curve is given by \cite{Han2}
\begin{equation}
{\Delta (m_{\nu _{2}}-m_{\nu _{1}})} _{DIA}=-2.5\log \left[ \left(
\frac{A_{\nu _{2}}(t)-1}{A_{\nu _{1}}(t)-1}\right) \left(
\frac{A_{\nu _{2}}(t_{{\rm ref}})-1}{A_{\nu _{1}}(t_{{\rm
ref}})-1} \right) ^{-1} \right] .\label{14}
\end{equation}
The advantage of measuring this curve, instead of that given by
Eq. (\ref{12}), is that it does not depend on the blending
parameters $f_{\nu,i}$ (equivalently, $B_{\nu,i}$). We shall
choose $t_{{\rm ref}}$ from the condition \be {\Delta (m_{\nu
_{2}}-m_{\nu _{1}})} _{DIA}=0,\ee when the reference star suffers
no amplification. Basically, $t_{{\rm ref}} \rightarrow \infty$.
Again, we shall fix our attention to the U and I bands of a
K-giant source star with dimensionless radius $R_{*}=0.1$ and
$T_{{\rm eff}}=4750K$. The results for positive and negative
lensing with different impact
parameters $k=b_{0}/r_{*}=B_{0}/R_{*}$ are shown in Figure 3.\\

Even when the DIA colour curve can have a different form when
compared with that produced only by limb-darkening, they both
depend on the same parameters, $A_{\nu _{1}}$ and $A_{\nu _{2}}$.
Hence, the same information can be extracted from both curves, but
with significantly reduced uncertainties in the DIA case, because
of the absence of blending. \\

\begin{figure}
\begin{center}
\includegraphics[width=8cm,height=9cm]{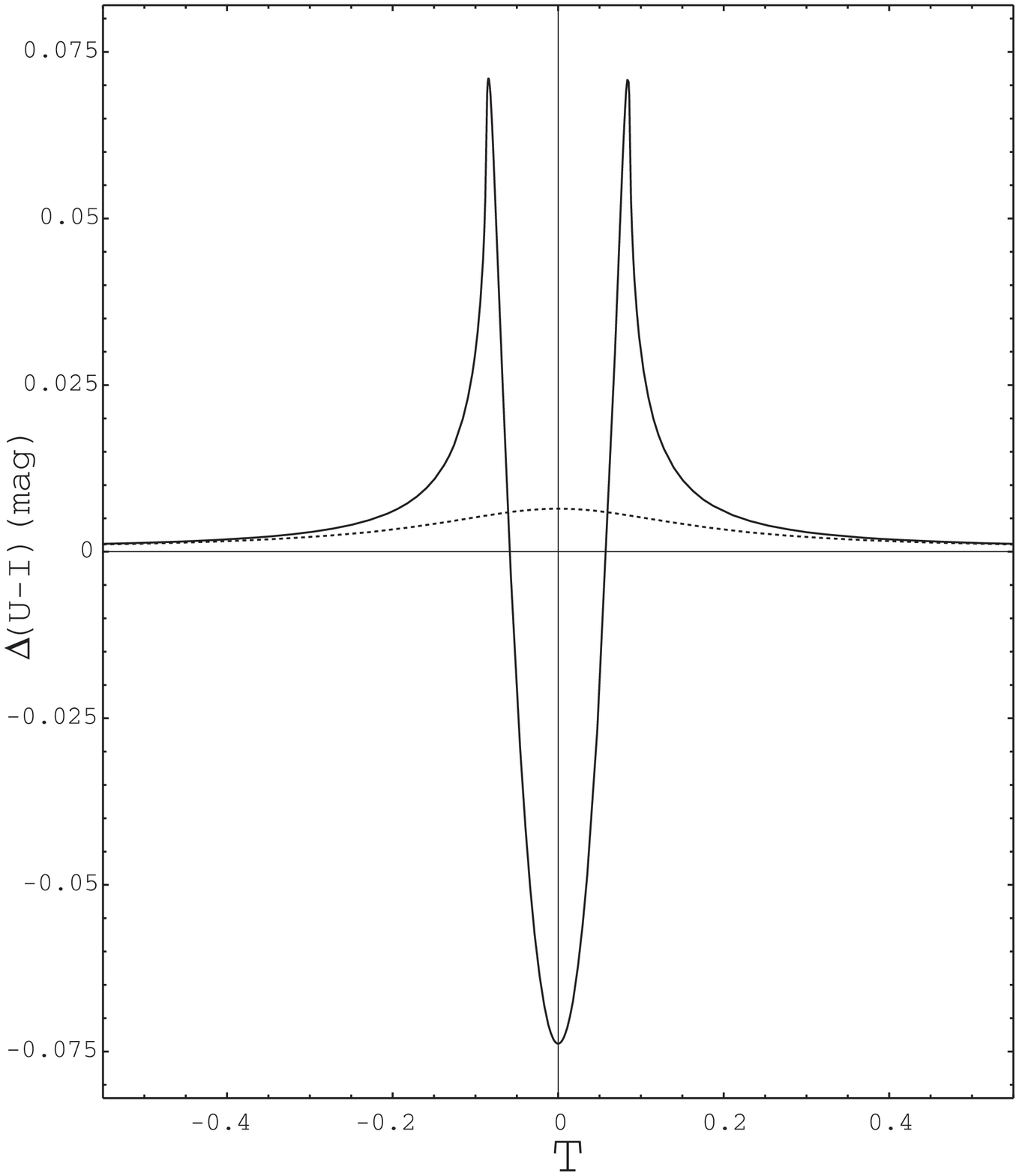}
\hspace{0.3cm}
\includegraphics[width=8cm,height=9cm]{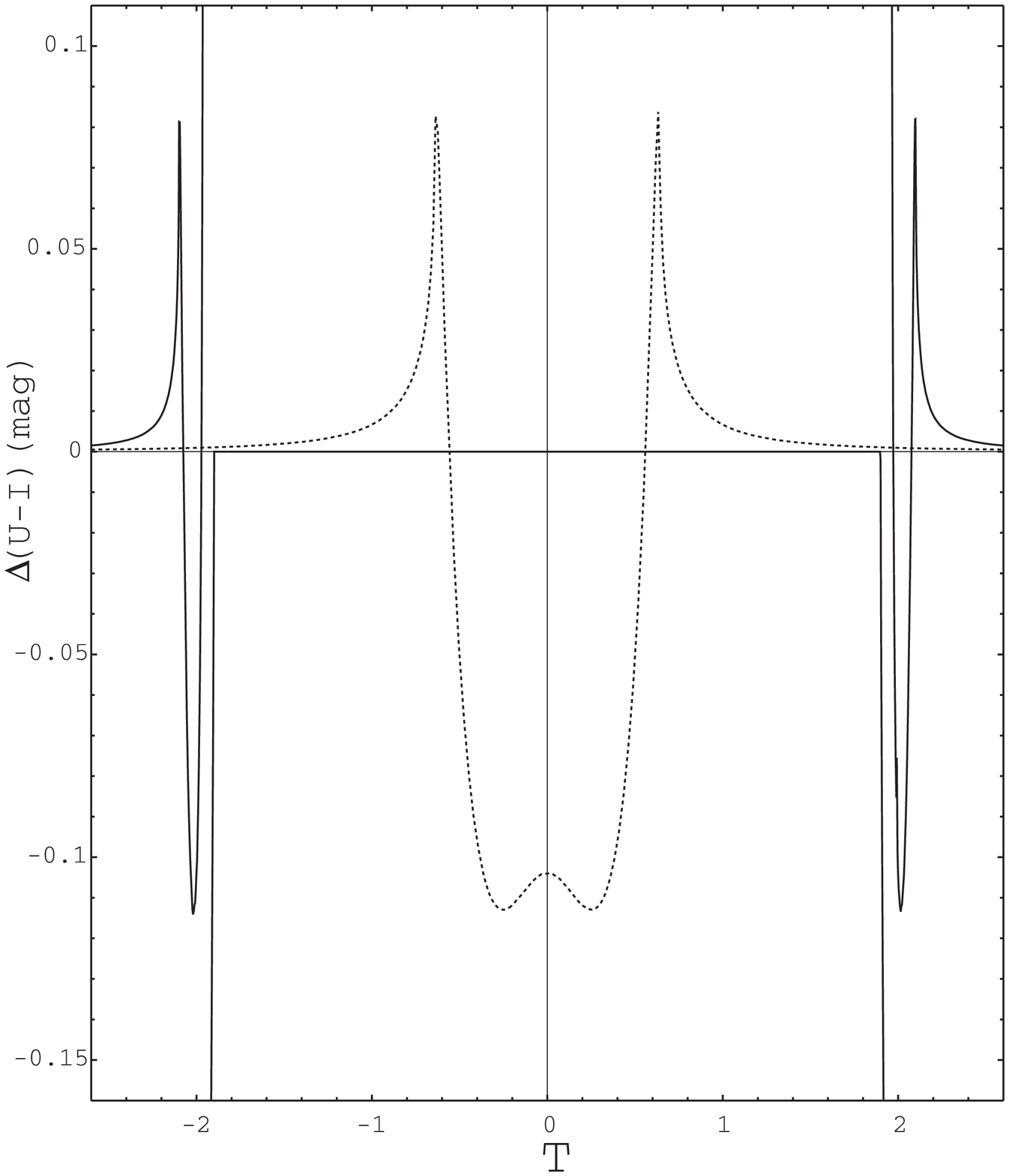}\\
 \caption{DIA colour curves. Left: Positive lensing, impact
parameter $k=b_{0}/r_{*}=0.5$ (solid curve) and $k=2$ (dash
curve). Right: Negative lensing,
$k=0.5$ (solid curve) and $k=20$
(dash curve). The
 source star has dimensionless radius $R_{*}=0.1$. \label{3}}
\end{center}
\end{figure}

\begin{figure}
\begin{center}
\includegraphics[width=8cm,height=9cm]{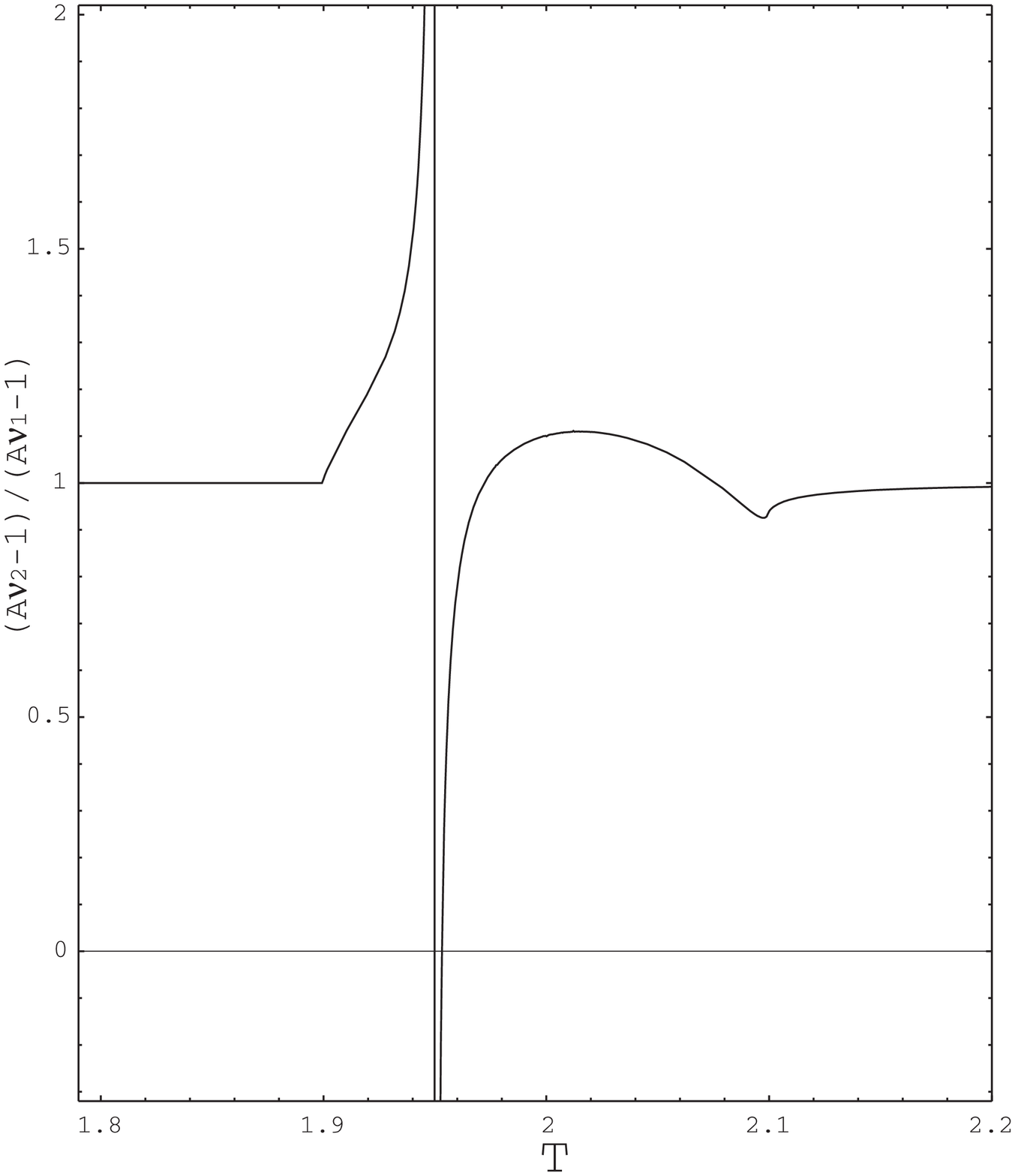}
\includegraphics[width=8cm,height=9cm]{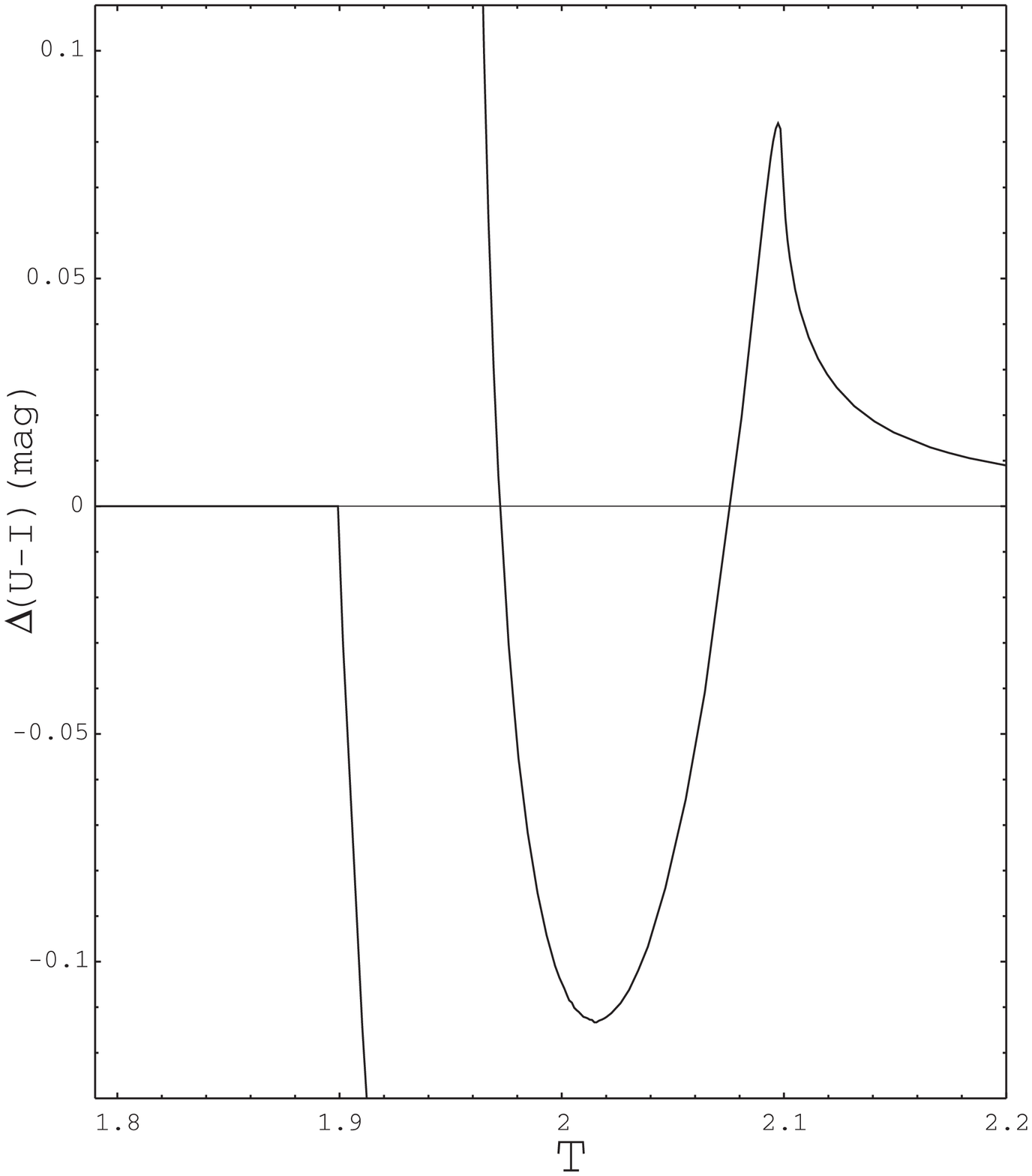}
 \caption{Left: Example of the partial evolution of the ratio $(A_{\nu _{2}}-1)/(A_{\nu
_{1}}-1)$ in the case of negative mass lensing. Right: DIA colour
curve in the same temporal interval. Impact parameter is $k=0.5$,
and $R_*=0.1$, for both figures. \label{f8} }
\end{center}
\end{figure}

It is interesting to directly compare, then, the DIA colour curve
just presented with the limb-darkening photometric curve presented
in Figure 3b of Ref. \cite{eiroa-croma}, or here in the right
panels of Figure \ref{2}, dash lines. The analytical difference
between both colour curves reduces itself to the replacement  \be
\frac{A_{\nu _{2}}}{A_{\nu _{1}}} \rightarrow \frac{A_{\nu
_{2}}-1}{A_{\nu _{1}}-1}\ee within the logarithm function used in
the magnitude definition. This apparently simple change has,
however, large implications for the negative mass colour curve
when $k < 20 \;\; ( b_0 < 2 R_E )$. Particularly, when either
$A_{\nu _{1}}$ or $A_{\nu _{2}}$ are less than 1, but not both,
the ratio $(A_{\nu _{2}}-1)/(A_{\nu _{1}}-1)$ is less than zero,
yielding a not defined colour change. This happens just before the
umbra, when large variations in the amplification suddenly occur
at slightly different times for different frequencies, this being
the reason of the apparent extra cusp in the DIA colour curve. We
show the behaviour of the ratio $(A_{\nu _{2}}-1)/(A_{\nu
_{1}}-1)$ for our two particular
frequencies in Figure \ref{f8}.\\

 Interestingly, the positive mass
DIA curve is completely similar to the photometric one, since
there is no time at which $A_{\nu _{1}}-1$ and $A_{\nu _{2}}-1$
have a different sign.\\

\begin{figure}
\begin{center}
\includegraphics[width=8cm,height=9.85cm]{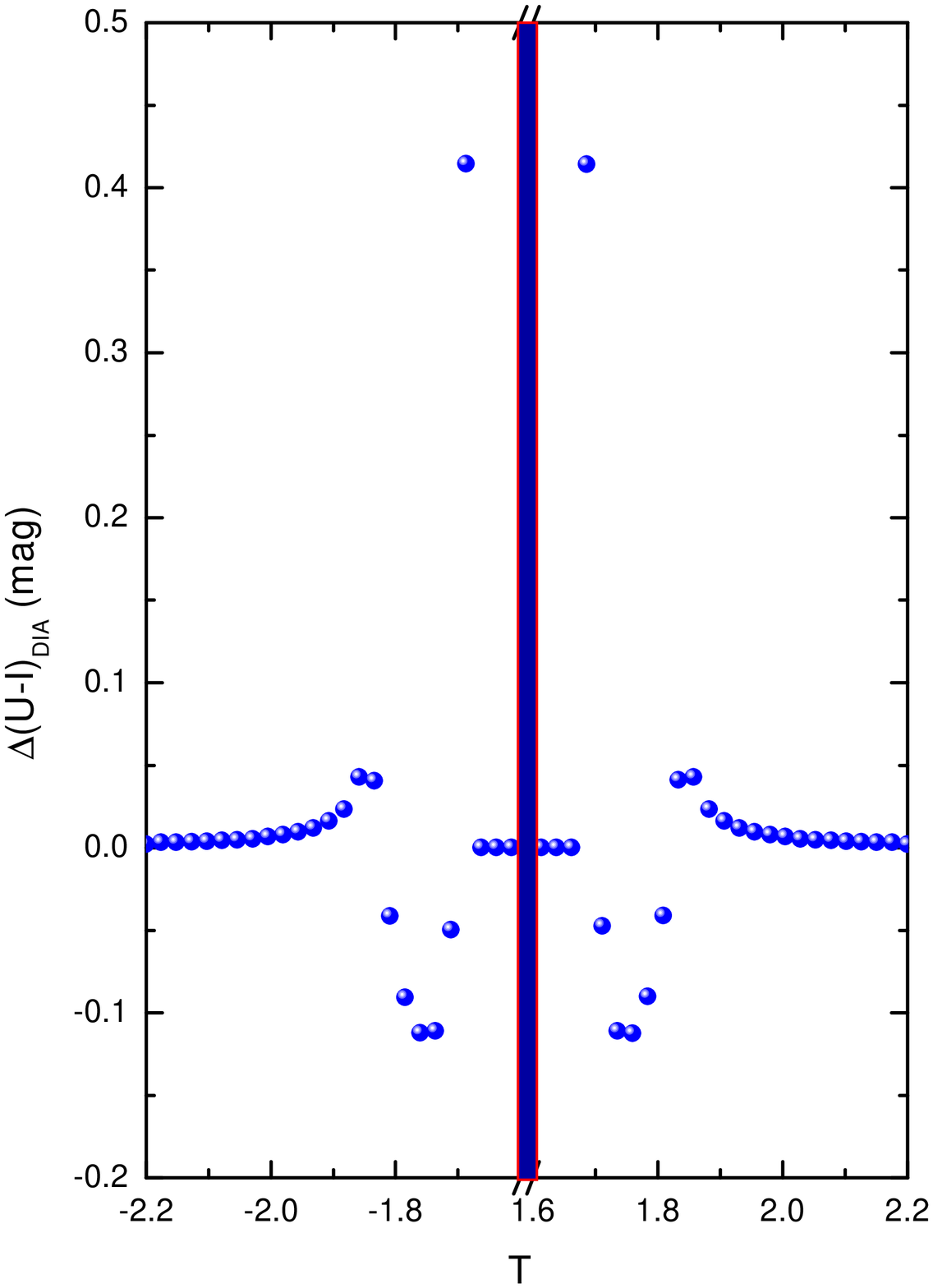}
\hspace{0.3cm}
\includegraphics[width=8cm,height=10cm]{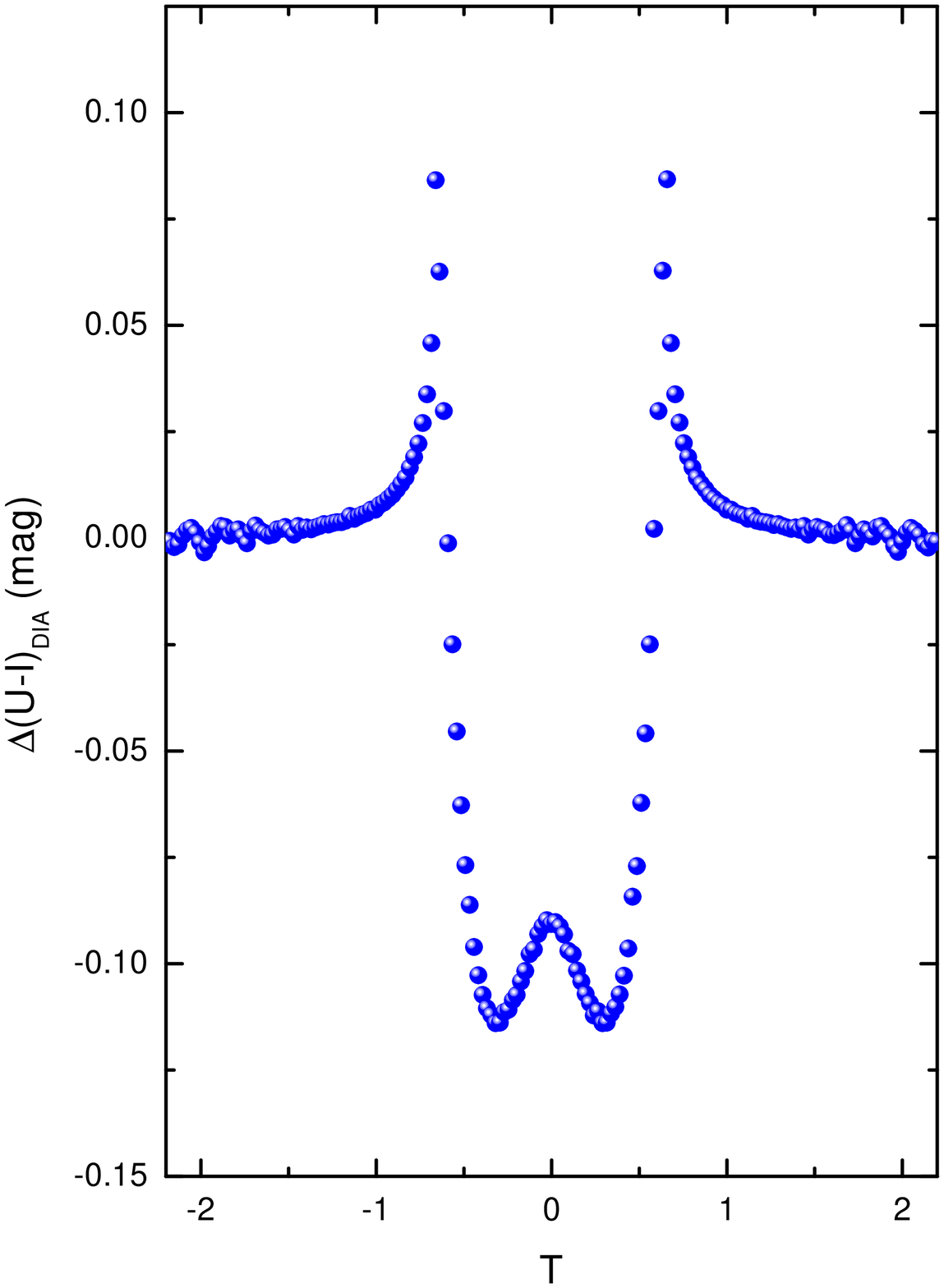}
 \caption{Negative mass DIA colour curves obtained with a numerical code, by
 binning the time interval for an impact parameter $k=10$ (left) and $20$ (right).
The left hand side presents a break on the
 $x$-axis of the colour curve, where the umbra is located,
 in order to provide extra details.
The
 source star has dimensionless radius $R_{*}=0.1$.
 \label{4}}
\end{center}
\end{figure}

The behavior of the negative DIA colour curve deserves further
study. In order to explore exactly the form of the curve that
could actually be measured, we would need to implement a numerical
code with a given binning in time (corresponding to a given
integration time of a telescope). If one of the cusps in the
colour curve is produced only by a single point, we might lose it
in the binning process, but we shall shed some light on the
behavior that could actually be observed. We have then adapted the
numerical code used in Refs. \cite{Han2} to the case of negative
mass lenses. Figure \ref{4} shows two particular examples obtained
with this code. These curves show the qualitative expected
behavior in its full extent.

\section{Measuring the DIA colour curve}

Although the colour changes are usually small, they can be
measured within current technological limitations. Following Han
et al. \cite{Han2}, we write the uncertainty in the determined
source star flux as related to the signal-to-noise ratio by \be
\delta m_\nu = \frac{\delta F_{\nu , 0} / F_{\nu , 0}}{0.4 \ln 10}
= \frac{1.09}{S/N}. \ee Then, the uncertainty in the measured
colour is related as well to $S/N$ by \be \delta \left[ \Delta
\left(  m_{\nu_2}-m_{\mu_{\nu_1}} \right)_{DIA} \right] \sim \sqrt
2 \delta m_\nu \sim \frac{1.54}{S/N} . \ee If $S/N=10$, $\delta
\left[ \Delta \left(  m_{\nu_2}-m_{\mu_{\nu_1}} \right)_{DIA}
\right] \sim 0.15$. The signal measured from the substracted image
is proportional to the source flux variation, \be S \propto (A_\nu
-1) F_{0,\nu} t_{\rm exp},\ee where $t_{\rm exp}$ is the exposure
time. The noise comes from the lensed source as well as from the
the blended background stars \cite{Han2}, \be N \propto [A_\nu
F_{0,\nu} + B ]^{1/2} t_{exp}^{1/2},\ee where $B$ represents the
average total flux of unresolved stars within a seeing disc of
radius $\theta_{\rm seeing}$. Then, the signal to noise ratio is
given by \be S/N=(A_\nu -1) F_{0,\nu} \left( \frac{t_{\rm
exp}}{A_\nu F_{0,\nu} + B} \right)^{1/2}.\ee Since we want to
compare our error estimates with those corresponding to a positive
case, we shall assume {\it mutatis mutandis} all parameters used
in the discussion of the latter in Section 5 of Ref. \cite{Han2}.
\\

Let us first take the source size as 0.07 Einstein radii, and the
Einstein time scale as 67.5/2 days \cite{Han2}. The lensed source
is a K-star with I=14.05 mag. Observations are assumed to be
carried with a 1m-telescope with a CCD camera that can detect 12
photons per second for a I=20 mag star. The exposure, $t_{\rm
exp}$, is considered variable so as to allow for the measured
signal to be $\sim 4 \; 10^4$ photons, which is in the range of
the linear regime response in modern CCD cameras. Actually, \be
 t_{exp} = \frac{ 4 \; 10^4 {\rm photons}}{A_\nu
F_{0,\nu} }, \ee and so it will be different for each given
magnification. The estimation of $B$ is done by assuming that
blended light comes from stars fainter (i.e. with greater
magnitudes) than the crowding limit, set when the stellar number
density reaches $\sim 10^6$ stars deg$^{-2}$. This number density
corresponds to $I\sim 18.1$ \cite{Han2}. The background flux is
normalized for stars in the seeing disc with $\theta_{seeing}=2$
arcsec. In the case of a positive lens, the exposure time required
to achieve the requested flux of $10^4$ photons is only about some
seconds, and this happens due to the huge magnifications that the
lensing produces (up to 20 times around $t=0$).\\

\begin{figure}
\begin{center}
\includegraphics[width=9cm,height=10cm]{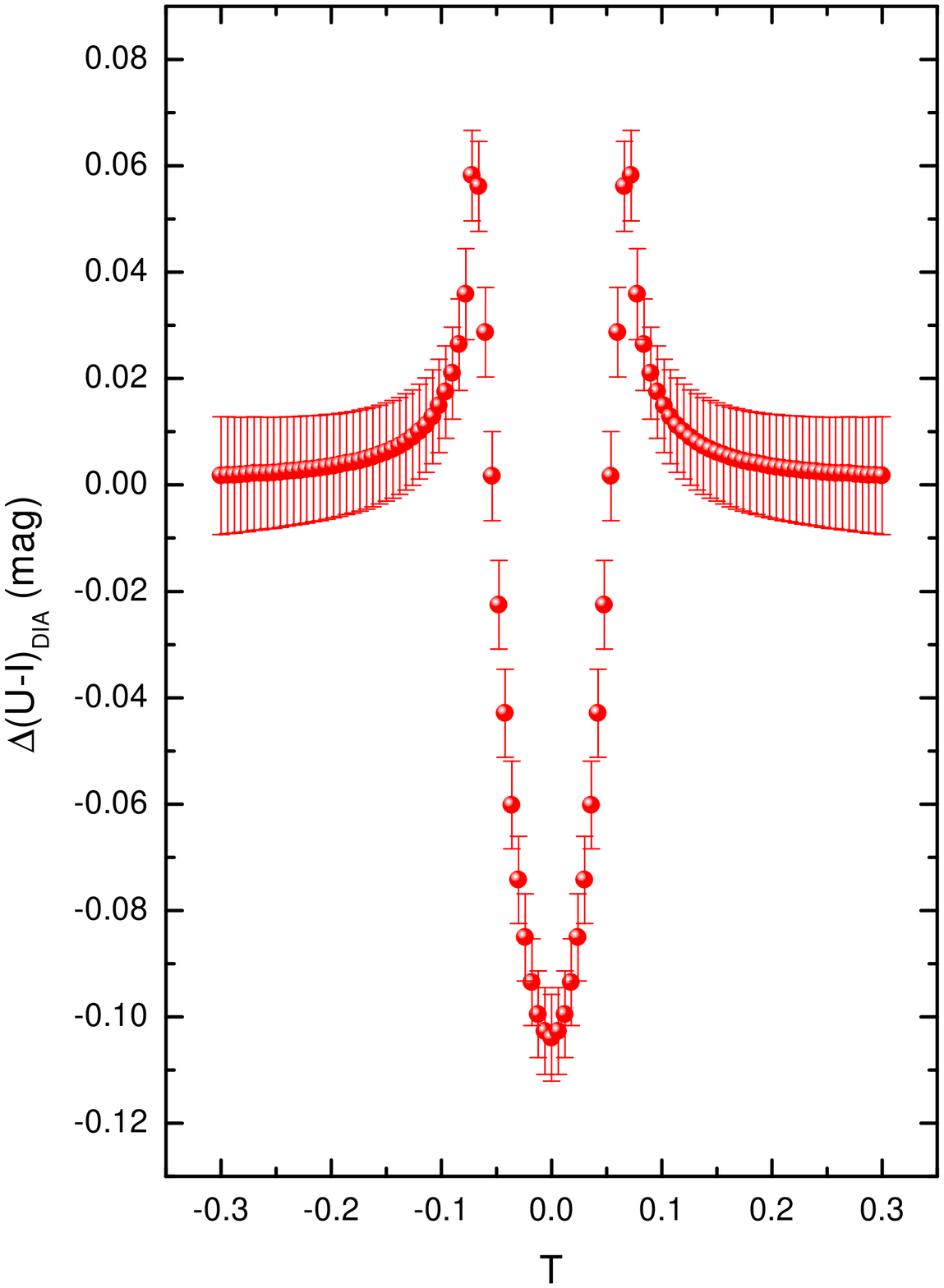}\hspace{-1.2cm}
\includegraphics[width=9cm,height=10cm]{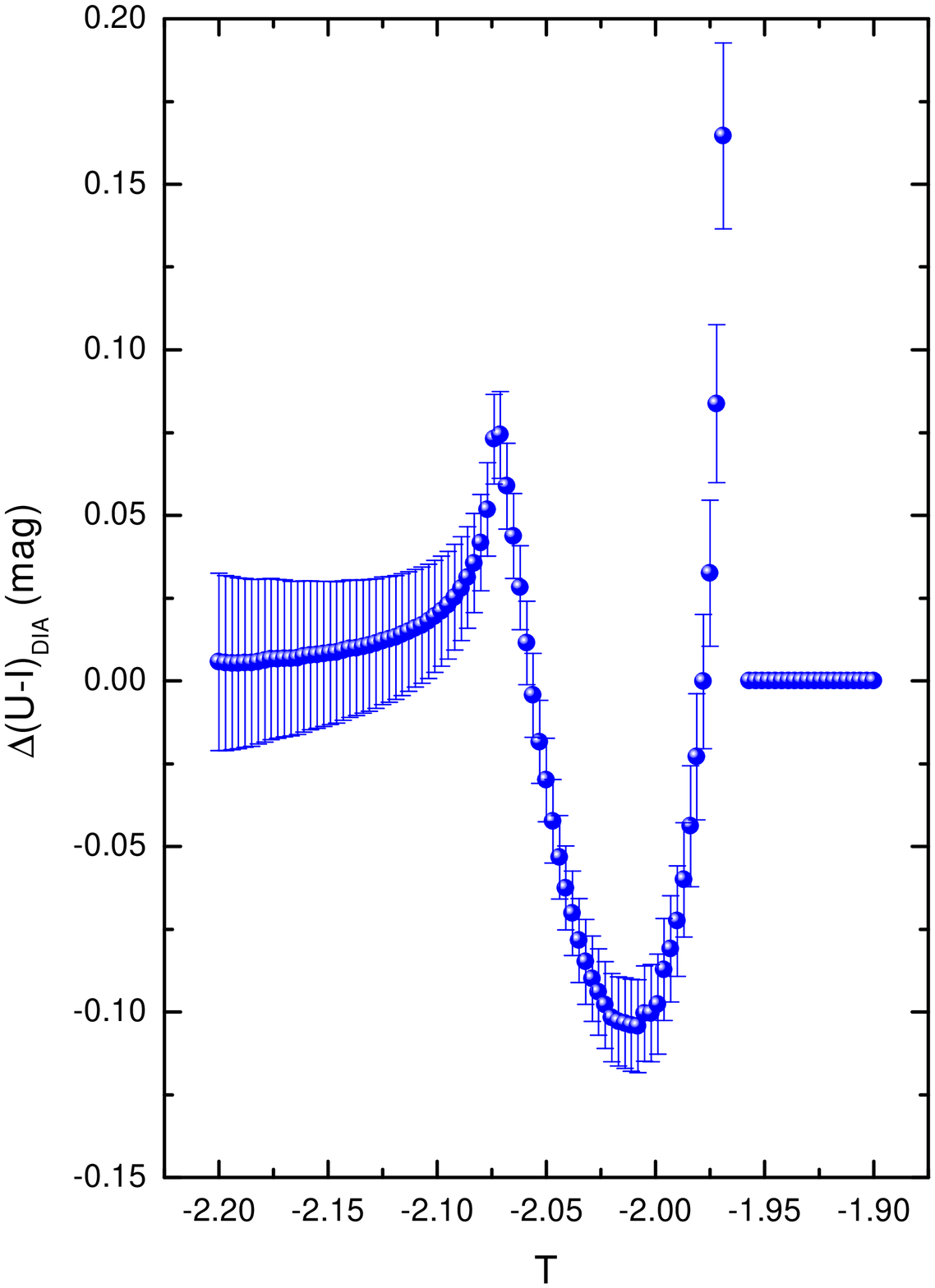}
 \caption{Left: Error estimate in the DIA colour curve for a
 positive lens, impact parameter $k=0.9$ and a source
 star with radius $R=0.07R_E$. Other parameters are discussed in
the text. Right: Example of the error estimate in the
 partial evolution (left of the umbra)
 of the colour curve for a negative lens. Lensing parameters are
 the same as in the Left panel.
  \label{error-1} }
\end{center}
\end{figure}

\begin{figure}
\begin{center}
\includegraphics[width=9cm,height=10cm]{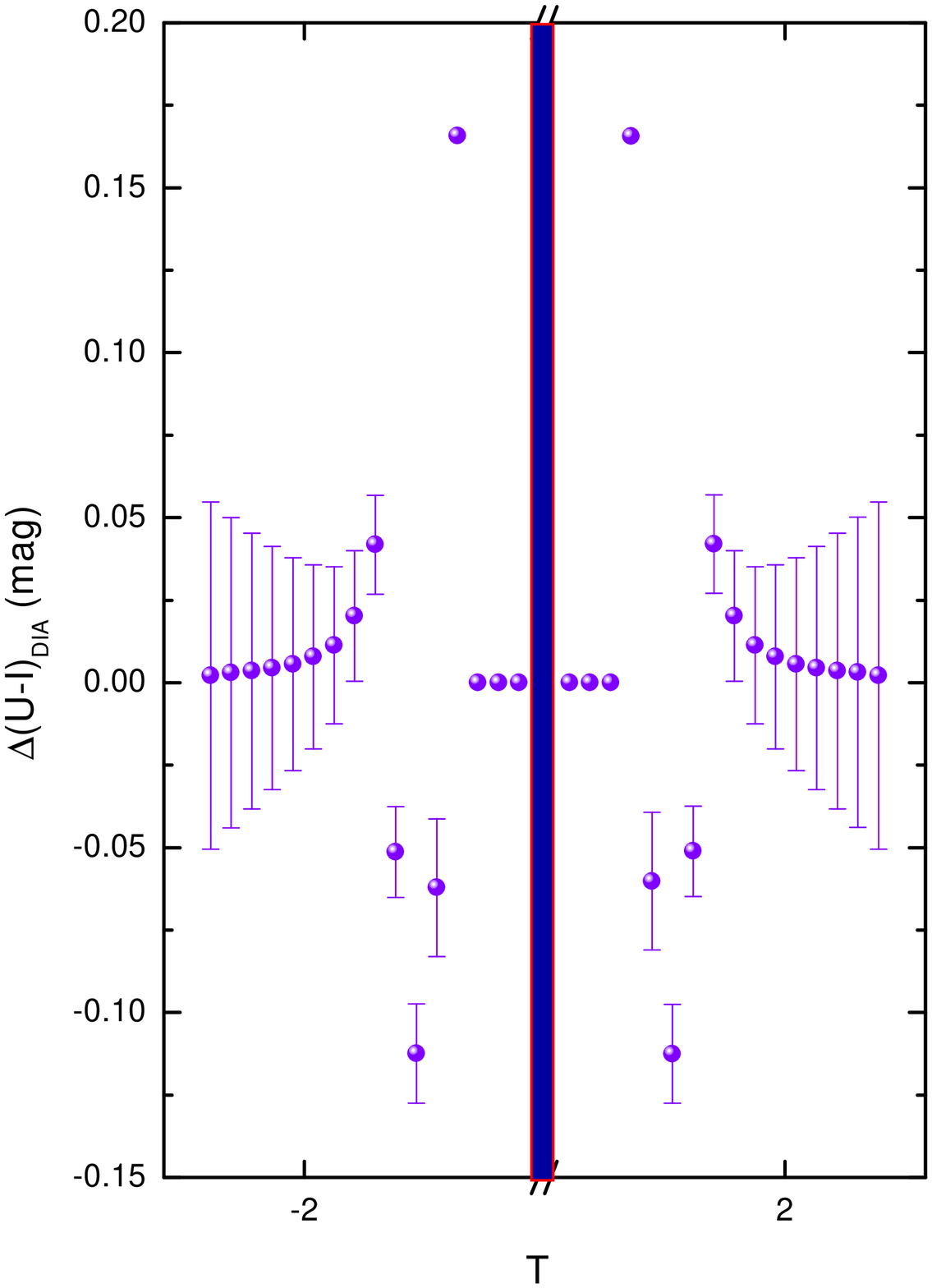}\hspace{-1.2cm}
\includegraphics[width=9cm,height=10cm]{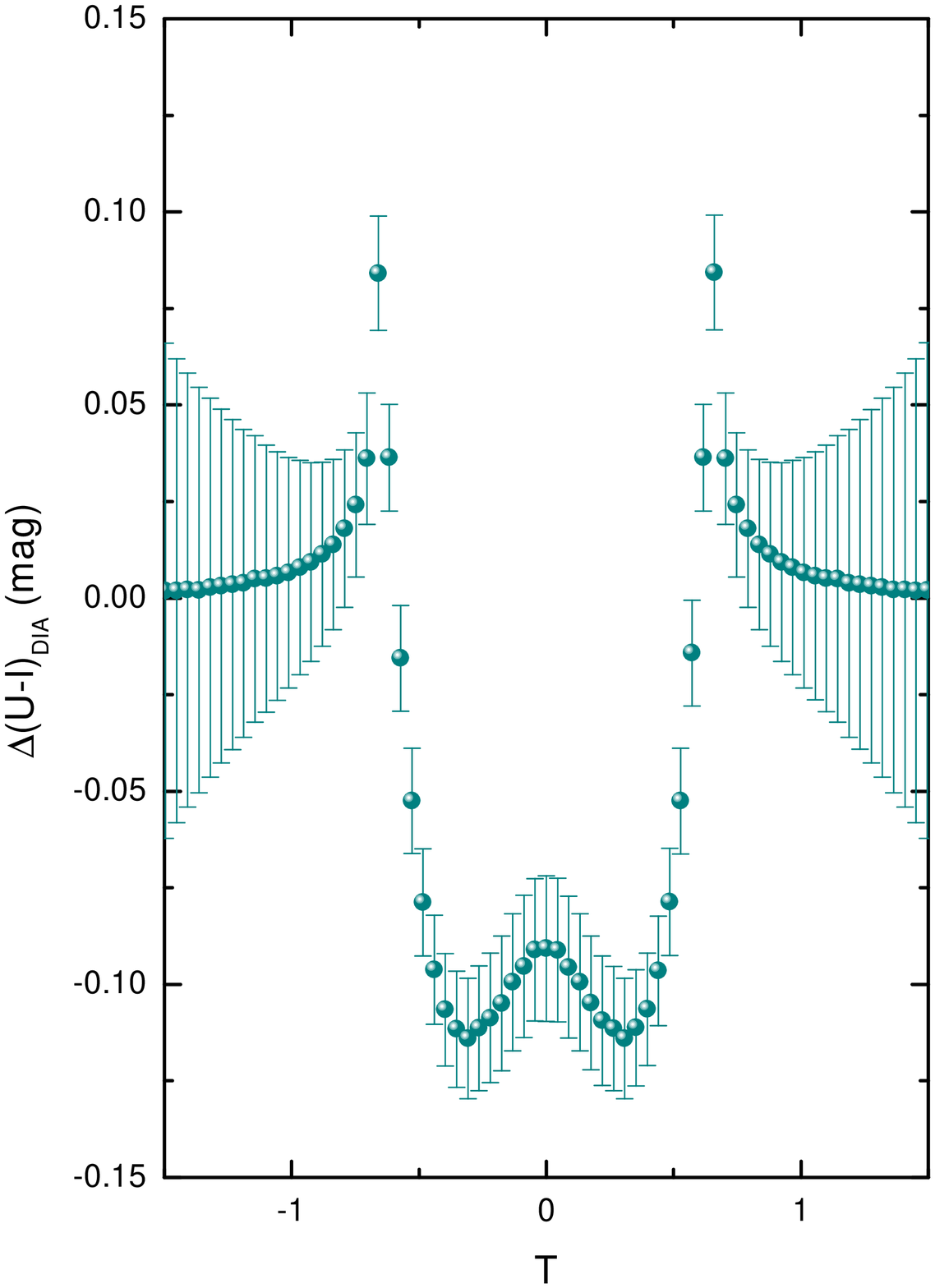}
 \caption{Error estimate in the DIA colour curve for
 negative lenses, impact parameter $k=10$ (again showing a break,
 left) and 20 (right), and a source
 star with radius $R=0.1R_E$.
  \label{error-2} }
\end{center}
\end{figure}

In the negative mass lensing situation, the overall presence of
the umbra dominates part of the error estimation as well. In
particular, for magnifications less than 1, the $S/N$ is not well
defined, since it becomes negative. But this happens just before
the umbra, for only one point in the binned plot, and do not
affect the correct estimate of the previously rising curve (on the
left of the umbra, for instance). In addition, there is no sense
in assigning an error to an absent signal, the umbra. We find that
$S/N$ for the negative case can be around 80, with exposures times
slightly larger than in the positive mass case, of about 8--10 s.
This difference is produced by generically lower values for the
magnification, which is of order 1, instead of the range 10--20
reached in the positive mass situation. In Figure \ref{error-1} we
show the case of $k=0.9$.  Note that the natural scale for
microlensing, the Einstein time scale, represents half the
physical time spread in the $x$-axis of the left panel in Figure
\ref{error-1}. Then, the negative mass lens has a longer time
evolution, since the particular peak we are showing happens
already in a time scale for which almost the complete microlensing
event occurs in the positive mass case. In Figure \ref{error-2} we
show the cases of impact parameters $k=10$ and $k=20$, for which
we have previously investigated the colour curve. Interestingly,
due to small values of the amplification for the earliest or the
latest times, the error significantly increases in these regions.
This can be noticed particularly on the right panel of Figure
\ref{error-2}. Overall, it is clear, then, that within current
observational capabilities we could be able to distinguish between
ordinary and exotic lenses, through the
analysis of gravitational microlensing chromatic effects.\\

At the moment, most of the microlensing experiments do not use the
DIA method in their data analysis. However this is already
beginning to change, see for instance Ref. \cite{og}, and will
become a common practice in the near future. If the microlensing
alert systems are adapted to take into account the possible colour
and light curves produced by negative mass lenses, we shall be in
position to make extensive searches --and to establish bounds on
the possible existence-- of wormhole-like objects.

\section{Concluding remarks}

All theoretical constructs thought to represent features of the
real world should be queried through experimental or observational
tests. This process is fundamental for science. In this paper, we
have expanded the formalism for wormhole-like gravitational
microlensing of extended sources by including the analysis of the
effects of blending. Having so constructed a complete colour
curve, taking into account the effects of limb darkening as well,
we analyzed the possibilities for an actual detection of
chromaticity effects.\\

Struts of negative masses, if they exist at all, will be detected
through the effects they produce upon the light coming from
distant sources. If a consistent lensing survey yields a negative
result, we could then set empirical constraints from a statistical
point of view to the amount of negative mass in the universe.

\section*{Acknowledgments}

This work has been supported by Universidad de Buenos Aires
(UBACYT X-143, EFE), CONICET (DFT, and PIP 0430/98, GER), ANPCT
(PICT 98 No. 03-04881, GER), and Fundaci\'{o}n Antorchas (through
separates grants to GER and DFT). DFT is on leave from IAR and
especially acknowledge Prof. Cheongho Han for providing him with
the basic numerical codes used in Figs. 5, 6 and 7 of this paper.



\begin{thebibliography}{99}


\bibitem{our}D. F. Torres, G. E. Romero \& L. A. Anchordoqui,
Phys. Rev. D{\bf 58}, 123001 (1998); D. F. Torres, G. E. Romero \&
L. A. Anchordoqui, {\it (Honorable Mention, Gravity Foundation
Research Awards 1998)}, Mod. Phys. Lett. {\bf A13}, 1575 (1998);
M. Safonova, G. E. Romero \& D. F. Torres, Mod. Phys. Lett. {\bf
A16}, 153 (2001) [astro-ph/0104075]; L. A. Anchordoqui, G. E.
Romero, D. F. Torres \& I. Andruchow, Mod. Phys. Lett. {\bf A14},
791 (1999); G.E. Romero, D.F. Torres, L.A. Anchordoqui, I.
Adruchow, B. Link, Monthly Notices Royal Astron. Soc. {\bf 308},
799 (1999); M. Safonova, G. E. Romero \& D. F. Torres,
[gr-qc/0105070]; L. A. Anchordoqui, S. Capozziello, G. Lambiase \&
D. F. Torres, Mod. Phys. Lett. {\bf A15}, 2219 (2000).

\bibitem{motho}M. S. Morris \& K. S. Thorne, Am. J. Phys. {\bf 56}, 395 (1988).

\bibitem{wh}D. Hochberg \& M. Visser, Phys. Rev. Lett. {\bf 81},
746 (1998); Phys. Rev D{\bf 58}, 044021 (1998); Phys. Rev. D{\bf
56}, 4745 (1997); E. E. Flanagan \& R. M. Wald, Phys. Rev. D{\bf
54}, 6233 (1996); L. A. Anchordoqui, S. E. Perez Bergliaffa \& D.
F. Torres, Phys. Rev. D{\bf 55}, 5226 (1997); C. Barcel\'o \& M.
Visser, Phys. Lett. B{\bf 466}, 127 (1999); A. DeBenedictis \& A.
Class.Quant.Grav. 18, 1187 (2001); S. E. Perez Bergliaffa \& K. E.
Hibberd, Phys. Rev. D {\bf 62}, 044045 (2000); L. A. Anchordoqui
\& S. E. Perez Bergliaffa, Phys. Rev. D {\bf 62}, 067502 (2000);
D. Hochberg, A. Popov, \& S. V. Sushkov, Phys. Rev. Lett. {\bf
78}, 2050 (1997); S. Kim \& H. Lee, Phys. Lett. B {\bf 458}, 245
(1999); S. Krasnikov, Phys. Rev. D {\bf 62}, 084028 (2000).

\bibitem{eiroa-croma}E. F. Eiroa, G. E. Romero, \& D. F. Torres, Modern Physics
Letters {\bf A16}, 973 (2001).

\bibitem{cramer} J. G. Cramer, R. L. Forward, M. S. Morris, M. Visser, G. Benford, G. A.
Landis, Phys. Rev. D{\bf 51}, 3117 (1995).

\bibitem{Han}C. Han, S-H. Park, J-H. Jeong, Monthly Notices Royal Astron. Soc.
{\bf 316}, 97 (2000).

\bibitem{Han2}C. Han, \& S-H. Park, Monthly Notices Royal Astron. Soc.
{\bf 320}, 41 (2001).

\bibitem{og}P.R. Wozniak et al., {\it Difference image analysis of
the OGLE-II bulge data II: Microlensing events}, astro-ph/0106474.

\end{thebibliography}
\end{document}